\documentclass[aps,pra,preprint,superscriptaddress,11pt]{revtex4-1}

\usepackage{color}
\usepackage{xcolor}
\usepackage{graphicx}
\usepackage{amsmath}
\usepackage{amsfonts}
\usepackage{soul}
\usepackage[normalem]{ulem}
\usepackage{lipsum}
\usepackage{bibentry}
\usepackage{float}

\DeclareMathOperator\erf{erf}

\usepackage{silence}
\begin{document}

\title{Automating physical intuition in nonlinear fiber optics with unsupervised dominant balance search}

\author{Andrei V. Ermolaev}
\affiliation{Universit\'{e} de Franche-Comt\'{e}, Institut FEMTO-ST, CNRS UMR 6174, 25000 Besan\c{c}on, France}

\author{Christophe Finot}
\affiliation{Universit\'{e} de Bourgogne, Laboratoire Interdisciplinaire Carnot de Bourgogne, CNRS UMR 6303, 21078 Dijon, France}

\author{Go\"{e}ry Genty}
\affiliation{Photonics Laboratory, Tampere University, FI-33104 Tampere, Finland}

\author{John M. Dudley}
\affiliation{Universit\'{e} de Franche-Comt\'{e}, Institut FEMTO-ST, CNRS UMR 6174, 25000 Besan\c{c}on, France}

\begin{abstract}
Identifying the underlying processes that locally dominate physical interactions is the key to understanding nonlinear dynamics. Machine-learning techniques have
recently been shown to be highly promising in automating the search for dominant physics,  adding important insights that complement analytical methods and empirical intuition. Here we apply a fully unsupervised approach to the search for dominant balance during nonlinear and dispersive propagation in optical fiber, and show that we can algorithmically identify dominant interactions in cases of optical wavebreaking, soliton fission, dispersive wave generation, and Raman soliton emergence. We discuss how dominant balance manifests both in the temporal and spectral domains as a function of propagation distance.
\end{abstract}

\maketitle

\thispagestyle{empty}

Amongst the many applications of artificial intelligence in science \cite{Jordan-2015,Genty-2021}, the use of machine learning to physically interpret nonlinear dynamics is one of the most fascinating developments \cite{Brunton-2016,Brunton-2022,Floryan-2022}.  An area of particular interest has been the automated identification of dominant physical processes \cite{Callaham-2021a, Callaham-2021b,Kaiser-2022}, determining the governing physics of a system through computation rather than the usual approach of analysis in asymptotic limits \cite{Barenblatt-1996}, or intuition based on experience \cite{Callaham-2021a}. The overall aim here is to automatically (\emph{algorithmically}) identify subsets of terms of the governing differential equation model that locally dominate the physics at different stages of propagation, and this has been successfully applied to problems in hydrodynamics, aerodynamics, turbulence modelling, and pattern formation \cite{Callaham-2021a, Kaiser-2022}.

In the particular field of optics, Ref.~\cite{Callaham-2021a} described one application to supercontinuum generation \cite{Dudley-2006}, and an application to optical rogue wave growth and decay in the nonlinear Schr\"{o}dinger equation (NLSE) has also been reported \cite{Ermolaev-2023}.
These previous studies in optics, however, were not fully unsupervised, as they both involved a manual optimisation step which was essentially based on \emph{a priori} knowledge of the underlying physics. In this paper, we apply a fully unsupervised technique to two representative propagation scenarios in nonlinear fiber optics, and show that dominant balance can indeed be identified in a fully automated way.  In what follows, we first present a general description of the approach, and then consider illustrative examples of both normal and anomalous dispersion regime dynamics.  In addition to mapping dominant balance in the temporal evolution domain as in previous studies, we show how dominant balance maps appear in the corresponding spectral domain.

The search for dominant balance aims to find a subset of terms of a propagation model that locally dominate the dynamics at particular points in the evolution \cite{Callaham-2021a}. Finding these terms is equivalent to identifying the corresponding dominant physical processes. To understand this more clearly, we write a general evolution equation on the spatio-temporal domain $(\xi,\tau)$ as:
\begin{equation}
     \sum_{i=1}^{K} f_{k}(\psi, \psi_{\xi}, \psi_{\tau} ..., \psi^{2}, \psi\psi_{\xi}, \psi\psi_{\tau},..., \psi_{\xi\xi}, \psi_{\tau\tau}, ... ) = 0,
\label{eq: general form}
\end{equation}
where $K$ is the total number of terms in the equation, and the $f_{k}$ are various possible differential operators and other functions of $\psi(\xi, \tau)$. This implicit form of the propagation equation stresses the balance that is needed for the terms to satisfy the equality by summing to zero. ``Dominant balance'' is the situation when only a subset of the terms $f_{k}$ dominate (or approximately dominate) the equality, with the contributions from the other terms being comparatively negligible.

\begin{figure*}[ht]
\centering
\fbox{\includegraphics[width=1\linewidth]{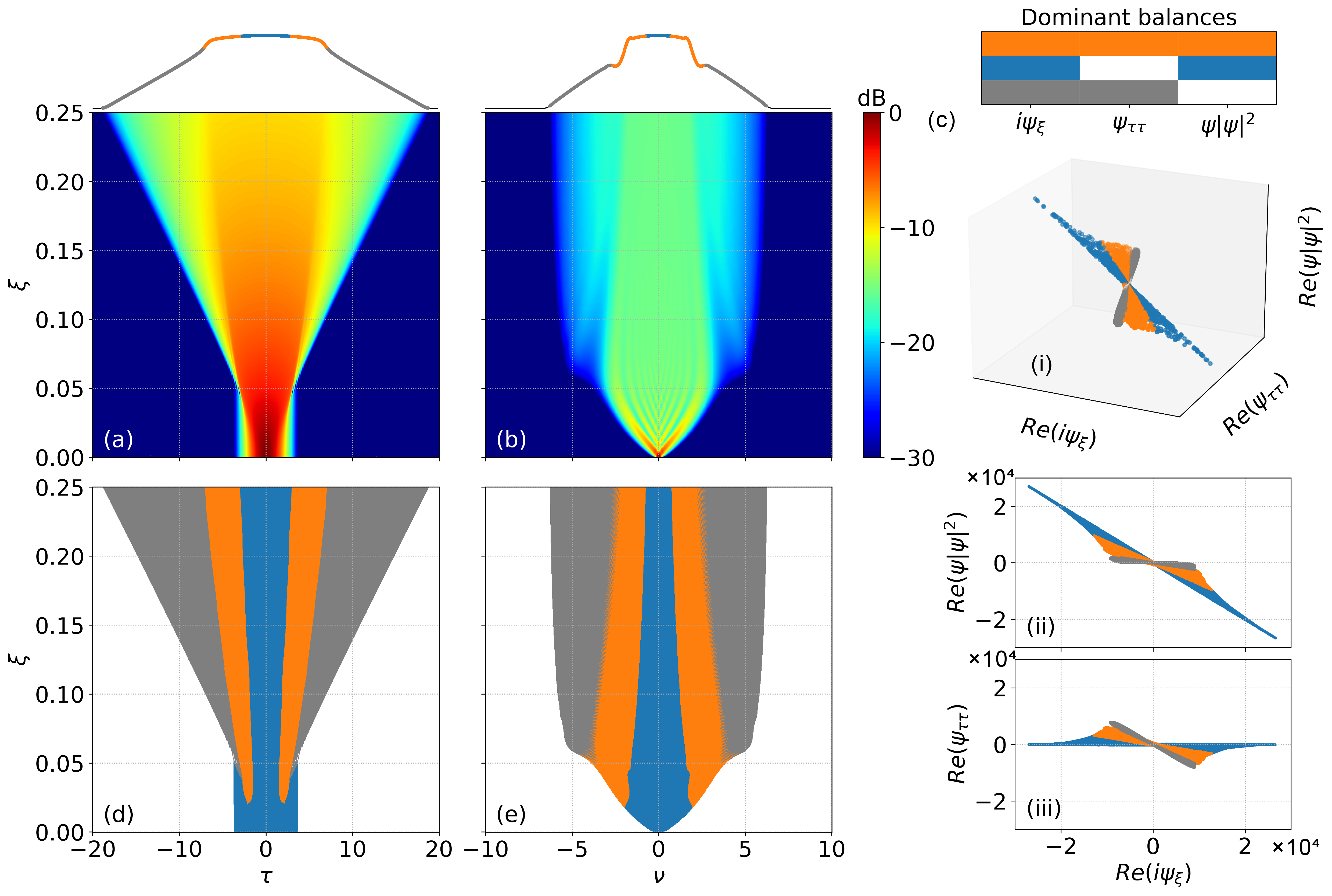}}

\caption{(a) temporal and (b) spectral intensity evolution plots for optical wavebreaking. Both plots are normalised to unity maximum intensity and use the logarithmic scale as shown. We plot above each figure the output temporal and spectral profiles on a linear scale. (c) illustrates clustering in: (i) the full equation space; and (ii,iii) for two projections. (d) and (e) plot the temporal and spectral dominant balance regions for comparison with the intensity evolution plots. The color key in (e) for the spectral plot maps to the Fourier transforms of the corresponding time-domain terms given in the legend.}
\end{figure*}

The algorithm begins by computing the evolution map $\psi(\xi,\tau)$ through numerical integration of the governing differential equation. The next step analyses $\psi(\xi,\tau)$ in its associated ``equation space,'' where each term of the governing equation is considered a coordinate axis. A Gaussian Mixture Model (GMM) is initially used to find correlations between different terms in equation space by dividing the points in the equation space into clusters with different covariances and means. This usually identifies many more clusters than there are potential physical combinations of terms, but as shown in \cite{Callaham-2021a}, sparse principal component analysis can be used to group  clusters with similar sparsity patterns (e.g. reduced variance in the same directions). These clusters can then be associated with a subset of terms that dominate the dynamics. However, principal component analysis requires manual optimisation of an $l_1$ regularization step to isolate physically-meaningful balances, which in practice requires \emph{a priori} knowledge of which balances may be present. Although still a useful technique \cite{Ermolaev-2023}, this is not fully unsupervised.

A recent alternative method replaces principal component analysis by a simple comparison of the magnitude of the different terms at each $(\xi,\tau)$ to determine when a particular subset of terms is dominant \cite{Kaiser-2022}. In this approach, dominance of a subset is associated with: (i) a maximised difference between the terms in the subset and the other terms; and (ii) a minimized difference between the terms within the subset itself. A metric to describe this condition can be readily defined (see Supplementary Information), and for each GMM cluster, this is computed for each combination of terms. The value of the metric indicates which physical balance is dominant in each cluster, and clusters with the same dominant balance are grouped together. This procedure also identifies cases where all terms in the governing equation contribute comparably. The $(\xi,\tau)$ points in each group are then assigned a color code to plot a dominant balance map for comparison with the evolution plot. This fully unsupervised approach eliminates need for advance knowledge of what results might be expected, and is the method we use here.

We first consider optical wavebreaking in the normal dispersion regime of fiber propagation. The governing NLSE in normalized form is: $i \, \psi_\xi - \psi_{\tau\tau} + |\psi|^2\psi = 0$.  Transformation to the common NLSE for dimensional field $A(z,t)$ \cite{Agrawal-2019} uses: $\psi = A \sqrt{\gamma L_\mathrm{D}}$, $\xi = z / L_\mathrm{D}$, and $\tau = t\,\sqrt{2} / T_{0}$. Here $\gamma$ and $\beta_2$ are the usual fiber nonlinearity and group-velocity dispersion, $P_0$ and $T_0$ are characteristic power and timescale, and length scale $L_\mathrm{D} = {T_0}^2/|\beta_2|$.  We consider a Gaussian input pulse $\psi(0,\tau) = N \, \exp{(-\tau^{2}/4)}$ with $N = T_0\sqrt{\gamma P_0 /|\beta_2|} = 30$. Figs 1(a) and (b) plot the computed temporal and spectral evolution of $|\psi(\xi,\tau)|^2$ and  $|\tilde{\Psi}(\xi,\nu)|^2$, where $\psi$ and $\tilde{\Psi}$ are Fourier transform pairs, and $\nu$ is dimensionless frequency. We also show output temporal and spectral profiles. These results show the expected characteristics of optical wavebreaking such as simultaneous temporal and spectral broadening, the development of a flat-top temporal profile, and spectral side-lobes \cite{Tomlinson-1985,Agrawal-2019}.

The results of the dominant balance search algorithm are shown in Fig. 1(c), where Fig. 1(c-i) plots the clusters in the coordinate space $ \{i\psi_\xi,\psi_{\tau \tau},\psi|\psi|^2\} $, whereas Figs. 1(c-ii) and 1(c-iii) show two projections as indicated.  For clarity we plot the real components, but similar results are found for the imaginary components. The color key corresponds to the three models that are found: where only dispersive and propagation terms contribute (gray); where only nonlinear and propagation terms contribute (blue) and where all three terms contribute (orange).

\begin{figure*}[ht]
\centering
\fbox{\includegraphics[width=\linewidth]{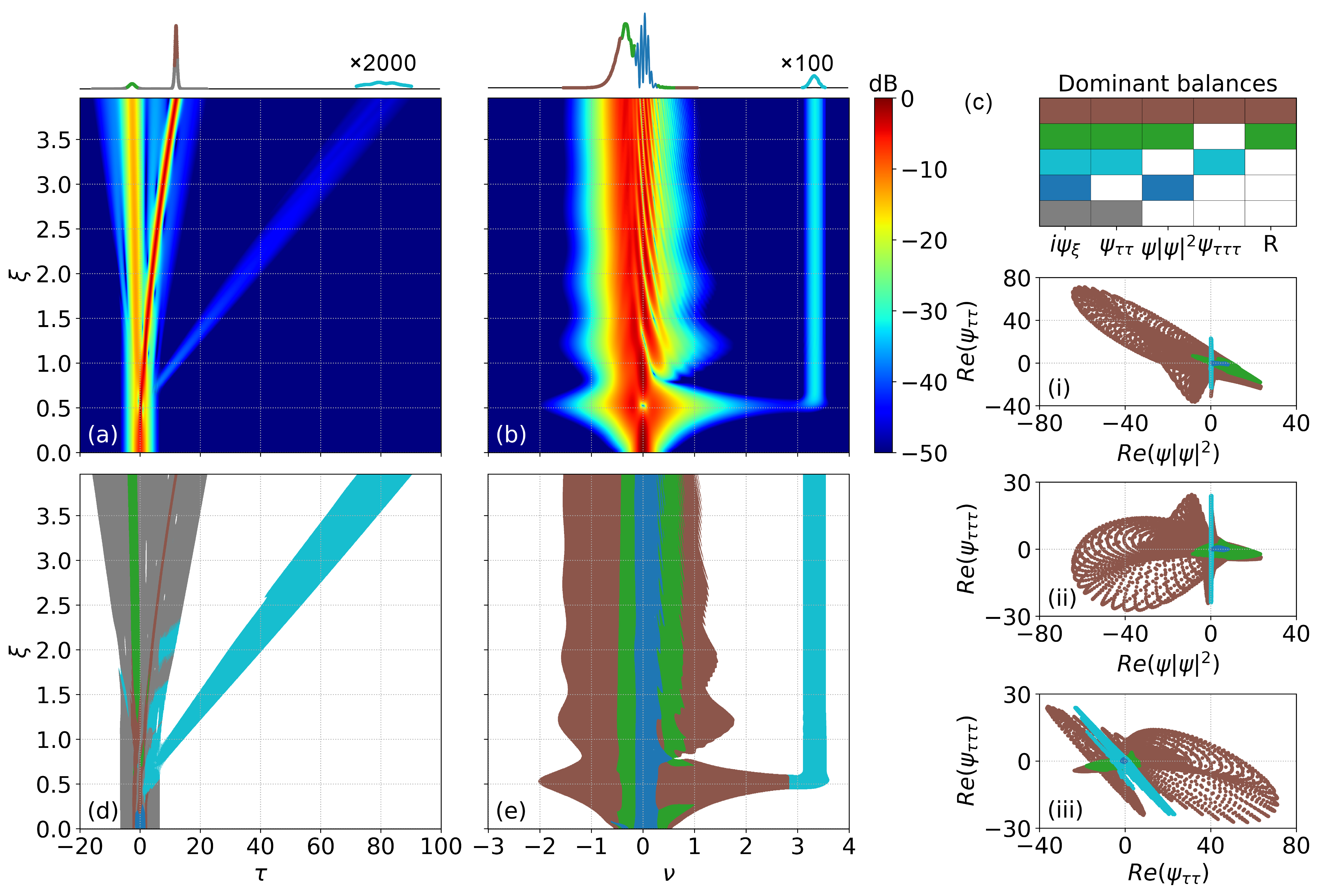}}
\caption{(a) temporal and (b) spectral intensity evolution plots for soliton fission. Both plots are normalised to unity maximum intensity and use the logarithmic scale as shown. We plot above each figure output temporal and spectral profiles on a linear scale. Because of their low intensity on a linear scale, the temporal and spectral signatures of the dispersive wave components (light blue) in the output profiles are magnified by the scale factors indicated. (c)(i-iii) show temporal clustering results in three projections. (d) and (e) plot temporal and spectral dominant balance regions for comparison with the intensity evolution plots. The color key in (e) for the spectral plot maps to the Fourier transforms of the corresponding time-domain terms in the legend.}
\end{figure*}

\newpage

These clustering plots highlight how the dominant balance of a subset of the NLSE terms has a geometric interpretation in equation space: two interacting terms are associated with a cluster falling on a line, while three interacting terms are associated with a cluster in a plane. For example, the projections in Fig 1(c-ii) show how the points assigned to the blue cluster (nonlinear and propagation terms) fall on a line with non-zero slope in $ \{i\psi_\xi,\psi|\psi|^2\} $, whilst the points assigned to the gray cluster (dispersive and propagation terms) fall on a line with non-zero slope in $ \{i\psi_\xi,\psi_{\tau \tau}\}$.  In contrast, the points assigned to the orange cluster (all NLSE terms) are distributed throughout equation space with no reduced variance relative to any axis. Figure 1(d) shows the color-coded clusters mapped back onto a segmented dominant balance plot in the time-domain. The same color code is used in the  line profiles above Fig.~1(a) and Fig.~1(b).

Although not considered in previous studies of dominant balance, nonlinear evolution equations can, of course, be equivalently expressed in the frequency-domain \cite{Francois-1991}. Indeed, for the dynamics shown in Fig.~1, it is straightforward to determine the corresponding dominant balance map in the spectral domain simply by implementing the algorithm on the computed Fourier transform of each term in the governing equation. This is shown in Fig.~1(e). The color key here maps to the Fourier transforms of the corresponding time-domain terms.

We now see how these results reveal the physics of optical wavebreaking. The temporal profiles in Fig.~1(a) and 1(d) show how the highest intensity regions are dominated by nonlinearity (dark blue), whereas the descending edges (orange) which become progressively steeper with propagation experience both dispersion and nonlinearity. As wavebreaking leads to the development of lower amplitude temporal wings (gray), these are associated with dominant dispersive propagation. The spectral plot in Fig.~1(e) shows how these balances are reflected in the frequency-domain, where we see the spectral sidelobes emerging at the onset of the dispersive dominance. Note that the visual similarity between these temporal and spectral correlations is expected in this case given the approximately linear time-frequency mapping of optical wavebreaking \cite{Finot-2008,Castello-2013,Heidt-2017}.

As a second example, we consider the anomalous dispersion regime dynamics of soliton fission in the presence of higher-order dispersion and Raman scattering. The generalized NLSE here is: $ i\psi_\xi +\psi_{\tau\tau} + i \delta \psi_{\tau\tau\tau} + |\psi|^2\psi +  \rho \psi\,(h_\mathrm{R}*|\psi|^2) = 0 $. The transformation to dimensional coordinates uses: $\psi = A \sqrt{\gamma L_\mathrm{D}}$, $\xi = z(1 - f_\mathrm{R})/L_{D}$, $\tau = t \sqrt{2(1 - f_\mathrm{R})}/T_{0}$. The third-order dispersion parameter $\delta = \sqrt{2(1 - f_{R})} \beta_{3}/ 3 T_{0} |\beta_{2}|$, and the Raman response function $h_\mathrm{R}$ is the standard two-timescale model of fused silica \cite{Agrawal-2019}. The operator $*$ represents convolution. We take $\delta = 0.05$ and $\rho = f_\mathrm{R}/(1-f_\mathrm{R})$ and Raman fraction $f_\mathrm{R} = 0.34$. We use hyperbolic secant initial conditions corresponding to a higher-order  $N = 2$ soliton, which with our normalisation corresponds to $\psi(0,\tau) = N \, \textrm{sech}[\tau/\sqrt{2(1 - f_\mathrm{R})}] $. These are conditions well-known to lead to soliton fission dynamics \cite{Dudley-2006}.

The results in Figs 2(a) and (b) again show the usual temporal and spectral evolution while Figs 2(c)-(e) show the results of the dominant balance analysis. Note that with more than 3 terms in the governing equation it is not possible to show a multi-dimensional cluster plot and Fig. 2(c) shows only three illustrative projections. The soliton fission process is very rich, and five different models are returned from the dominant balance algorithm: where only the second-order dispersive and propagation terms contribute (gray); where only the nonlinear and propagation terms contribute (dark blue); where both second- and third-order dispersive terms and propagation term contribute (light blue); where the three NLSE terms and the Raman term contribute (green); and where all terms contribute (brown).  Note that because of their low intensity when plotted on a linear scale, the temporal and spectral signatures of the dispersive wave components (light blue) in the output profiles above Figs 2(a) and (b) are scaled by the factors shown.

These results again illustrate the power of automated dominant balance analysis, as we can readily associate different elements of the temporal and spectral evolution with distinct combinations of physical processes.  For example, from Figs 2(d) and (e) we see clearly how the ejection of the dispersive wave at the onset of soliton fission around $\xi \sim 0.7$ is dominated by second and third-order dispersion (light blue) in both the temporal and spectral domains. This is of course fully consistent with the known phasematching condition for dispersive wave generation involving $\beta_2$ and $\beta_3$ \cite{Agrawal-2019}. We also see how the dominance of these two dispersive terms recurs around $\xi \sim 1.2$ and $\xi \sim 2$, with additional light blue regions apparent in the temporal field structure. This is associated with the well-known periodicity in the temporal and spectral evolution of the evolving soliton, and associated dispersive wave generation \cite{Cristiani-2004, Tran-2009}.

We also see how the central temporal structure consists of two localised soliton structures (green and brown) upon a broader dispersion-dominated background (gray). Here, dominant balance indicates that both soliton pulses experience the Raman effect, and the higher amplitude soliton also experiences third-order dispersion leading to the accelerating temporal trajectory. The narrow temporal duration of the accelerating soliton results in it dominating the spectral balance map (brown), although we can also see signatures associated with the lower-amplitude soliton (green). Of course, the basic soliton dynamics of this process have been extensively studied using simulations over a range of parameters \cite{Dudley-2006}, but the dominant balance analysis directly adds new insights such as the fact that the two soliton structures are not dominated by the same physical effects. We can see that the higher-amplitude soliton (brown) is dominated by all the effects, while the contribution of third-order dispersion is negligible for the lower-amplitude soliton (green).

There are several major conclusions to be drawn from this work. Firstly, the examples considered for the cases of optical wavebreaking and soliton fission (as well as the examples in the Supplementary Information) very clearly illustrate the power of dominant balance analysis to directly identify how different physical processes contribute to different stages of nonlinear and dispersive propagation.  We stress specifically that, in contrast to previous studies of dominant balance in optics \cite{Callaham-2021a,Ermolaev-2023}, our approach is completely unsupervised, opening the pathway to a fully automated method to yield ``intuitive'' understanding of the underlying physics.  Of course such a method does not replace human intuition completely, but it can provide very useful complementary information to stimulate new ways of thinking about seemingly well-known phenomena.   We can readily anticipate many interesting further applications of this technique, not only in nonlinear fiber optics, but in all optical systems where dynamics are described by differential equation models.  Moreover, the ability to identify local regions of a field dominated only by a subset of physical effects may open up new analytical studies using approximate perturbative methods. And finally, the analysis of propagation dynamics in terms of term-by-term contributions in the underlying equation space is a further interesting area of future study.

\vskip 2pt

\section* {Funding} Centre National de la Recherche Scientifque (MITI Ev\`{e}nements Rares 2022); Agence Nationale de la Recherche (ANR-15-IDEX-0003, ANR-17-EURE-0002, ANR-20-CE30-0004); Academy of Finland (318082, 320165 Flagship PREIN, 333949).

\section*{Acknowledgements} Sincere thanks are extended to Bryan Kaiser for discussions on the algorithm implementation.


\newpage

\section*{Supplementary Information}

\section{Summary}
This additional material provides further details of methodology, and several more
examples of unsupervised dominant balance search during nonlinear and dispersive propagation in optical fibre. Complementing the cases of optical wavebreaking and soliton fission in the primary manuscript above, we show that we can algorithmically identify dominant interactions in further cases of Riemann wave evolution in the normal dispersion regime; and perturbed soliton propagation and third-order dispersion induced soliton fission in the anomalous dispersion regime.

\section{Methodology}

The methodology for determining dominant balance models for a physical system at a particular stage of propagation aims at searching for a subset of terms of a more broadly applicable propagation model that dominate the dynamics locally.  In what follows, we describe the procedure for the governing equation expressed in terms of $(\xi,\tau)$, but the procedure is identical for a model in the  frequency-domain expressed in terms of $(\xi,\nu)$.
\\
\\
\textbf{Equation space representation.}
The general equation describing the field evolution in the spatio-temporal domain $(\xi,\tau)$ can be written in the following form:
\begin{equation}
     \sum_{k=1}^{K} f_{k}(\psi, \psi_{\xi}, \psi_{\tau} ..., \psi^{2}, \psi\psi_{\xi}, \psi\psi_{\tau},..., \psi_{\xi\xi}, \psi_{\tau\tau}, ... ) = 0,
\label{eq: general form1}
\end{equation}
\noindent where $K$ is the total number of terms in the governing equation, $f_{k}$ are differential operators and mathematical functions of the field $\psi({\xi, \tau})$. In most cases, the spatio-temporal field map $\psi({\xi, \tau})$ can be obtained by solving the equation~\ref{eq: general form1} numerically (in some cases analytically), but experimental data allowing access to the full field characterization can also be used.``Dominant balance'' refers here to the scenario when, on some subregion of the spatio-temporal domain $(\xi,\tau)$, the dynamics is approximately governed by only a subset of $S$ terms of the original equation~\ref{eq: general form1}. From another perspective, that implies that the contribution from the subset of $D = K - S$ residual terms is small or negligible.

\newpage

The starting point is the evolution of $\psi({\xi, \tau})$ discretized as $\psi({\xi_{n}, \tau_{m}})$ with indices $n \in [1,N]$ and $m \in [1,M]$, where $N$ and $M$ are the number of points in $\xi$ and $\tau$ respectively. This is written as:
\begin{equation}
\begin{pmatrix}
\psi({\xi_{1}, \tau_{1}}) & \psi({\xi_{1}, \tau_{2}}) & \dots & \psi({\xi_{1}, \tau_{M}})\\
\psi({\xi_{2}, \tau_{1}}) & \psi({\xi_{2}, \tau_{2}}) & \dots & \psi({\xi_{2}, \tau_{M}})\\
\vdots & \vdots & \ddots & \vdots \\
\psi({\xi_{N}, \tau_{1}}) & \psi({\xi_{N}, \tau_{2}}) & \dots & \psi({\xi_{N}, \tau_{M}})\\
\end{pmatrix}
\label{eq: field}
\end{equation}
With access to the field $\psi({\xi_{n}, \tau_{m}})$, the different terms $f_k$ in the equation space can be estimated numerically across the whole spatio-temporal domain. The evolution of the field $\psi(\xi, \tau)$ is thus fully represented by a $(NM \times K)$ matrix $\mathbf{\Lambda}$, each column of which constitutes a dynamical point in the $K-$dimensional equation space. This matrix is defined as follows:
\begin{equation}
\mathbf{\Lambda} =
\begin{pmatrix}
f_{1}[\psi({\xi_{1}, \tau_{1}})] & f_{1}[\psi({\xi_{1}, \tau_{2}})] & \dots & f_{1}[\psi({\xi_{1}, \tau_{M}})] & f_{1}[\psi({\xi_{2}, \tau_{1}})] & \dots & f_{1}[\psi({\xi_{N}, \tau_{M}})]\\
f_{2}[\psi({\xi_{1}, \tau_{1}})] & f_{2}[\psi({\xi_{1}, \tau_{2}})] & \dots & f_{2}[\psi({\xi_{1}, \tau_{M}})] & f_{2}[\psi({\xi_{2}, \tau_{1}})] & \dots & f_{2}[\psi({\xi_{N}, \tau_{M}})]\\
\vdots & \vdots & \ddots & \vdots & \vdots & \ddots & \vdots \\
f_{K}[\psi({\xi_{1}, \tau_{1}})] & f_{K}[\psi({\xi_{1}, \tau_{2}})] & \dots & f_{K}[\psi({\xi_{1}, \tau_{M}})] & f_{K}[\psi({\xi_{2}, \tau_{1}})] & \dots & f_{K}[\psi({\xi_{N}, \tau_{M}})]\\
\end{pmatrix},
\label{eq: equation space}
\end{equation}

As emphasised in~Ref.~\cite{Callaham-2021a}, ``dominant balance'' has a relatively simple geometric interpretation in the corresponding equation space: the ``dominant balance'' of $S$ terms manifests itself as a cluster of dynamical points that is restricted to $S$ directions of the full $K$-dimensional space. In other words, this cluster will have a significantly reduced variance with respect to other $K-S$ directions. For example, referring to Fig.~1(c) of the main manuscript for the simplest case of NLSE dynamics with only 3 terms $\{i\psi_\xi,\psi_{\tau \tau},\psi|\psi|^2\}$, it is evident how the clusters of dynamical points attributed to the dominance of the dispersion and propagation terms (gray), as well as to the dominance of the nonlinearity and propagation terms (blue) are approximately restricted to have the reduced variance with respect to the corresponding neglected terms. In contrast, the orange cluster representing the full NLSE dynamics is distributed throughout the equation space, not restricted to any of the three axes. Note that as in this case we are dealing with the complex-valued fields, Fig.1(c) displays only the real part of the $\mathbf{\Lambda}$ matrix, but similar cluster distributions are found for the imaginary components. \\

\newpage

\textbf{Unsupervised search for equation space clusters.}  The problem of identifying clusters in the equation space that represent different dynamical regimes can be addressed using various unsupervised clustering techniques. Among these, the use of K-means clustering~\cite{Sonnewald-2019}, DBSCAN~\cite{Kaiser-2022} and Gaussian Mixture Modeling (GMM)~\cite{Callaham-2021a} have recently been reported. Unsupervised cluster analysis using GMM can yield an initial partitioning of the equation space into distinct regions by training probabilistic models assuming that the data consist of a mixture of Gaussian distributions with different mean and covariance. In the context of the dominant balance search, the implementation of GMM clustering is preferred, since the learned covariance structure of GMM clusters can be useful in the physical interpretation of the results. The specific GMM algorithm used in this work is \texttt{GaussianMixture} from the Python \texttt{scikit-learn} package \cite{Pedragosa-2011} as implemented in Ref.~\cite{Callaham-2021b}. Even though in general case feature scaling is not required when performing the GMM clustering on the dataset, it is usually a good practice to use standardization in this case. In fact, the aforementioned implementation of the GMM uses the K-means-based strategy to set the initial starting values of the Expectation-Maximization algorithm and, thus, can be sensitive to the feature scaling. In order to better separate the equation space into the subsets of dynamical points, we use standardisation of the $\mathbf{\Lambda}$ matrix prior the GMM clustering. This process involves rescaling each feature such that it has standard deviation of 1 and mean of 0 (see \texttt{StandardScaler} from \texttt{sklearn.preprocessing}~\cite{Pedragosa-2011}). Note that this step is implemented only internal to the GMM algorithm; the cluster plots here and in the main manuscript plot the different terms based on the field variable $\psi$.

The GMM is applied directly to the real and imaginary parts of the standardized equation space matrix $\mathbf{\Lambda}$ to detect clusters that have significant variance along some axes, regardless of whether in the real or imaginary direction. The number of clusters to use as the input to the GMM procedure must be selected to be equal to or greater than the number of potential combinations of terms as governed by the structure of the governing differential equation \cite{Callaham-2021b,Kaiser-2022}. However, the precise value of this parameter is not critical because the next step in the algorithm can group different clusters and assign them to one of the potential dominant balances. For the results shown here, we used twice the maximum number of potential combinations of terms as input to the GMM procedure, and found this yielded excellent results.  We also note that we always include the propagation term in our cluster search which implies we are in a dynamically-evolving regime and not a steady-state. \\

\newpage

\textbf{Dominant Balance selection.} The last step of the method is to apply the combinatorial selection algorithm described in~\cite{Kaiser-2022} to automatically assign points in different clusters to particular dominant balances. It is this step that allows our algorithm to operate in a fully unsupervised manner, in contrast to the use of principal component analysis which requires manual hyperparameter optimisation \cite{Callaham-2021a}.  In particular, we describe the dominance of a subset as being associated with: (i) a maximised difference between the terms in the subset and the other terms (which we denote $\Gamma$); and (ii) a minimized difference between the terms within the subset itself (which we denote $\Omega$). To define these criteria using a computable metric, we introduce the normalised equation space matrix $\mathbf{\hat{\Lambda}}$, each column $\mathbf{\hat{\Lambda}}_{j}$ of which is normalised to the smallest nonzero value in the corresponding column:
\begin{equation}
     \mathbf{\hat{{\Lambda}}}_{j} = \frac{|\mathbf{\Lambda}_{j}|}{\min_{k \in \mathbf{F}}(|{\Lambda}_{jk}|)} \,\, ,
\label{eq: normalised equation space matrix}
\end{equation}
\noindent where $j \in [1,...,NM]$, $\mathbf{F} = \{1,...,K\}$, and $\min_{k \in \mathbf{F}}(|{\Lambda}_{jk}|) \neq 0$. This normalisation implies that each element of $\mathbf{\hat{{\Lambda}}}$ is greater or equal to 1. Next, for each of the potential dominant balances, we define the subset of the selected terms $\mathbf{s}_{j} \subseteq \mathbf{\hat{{\Lambda}}}_{j}$ of length $S$ and the subset of the residual terms $\mathbf{d}_{j} \subset \mathbf{\hat{{\Lambda}}}_{j}$ of length $D$ at every dynamical point.

\noindent The metric $M$ that describes the degree of dominance of particular terms is defined as follows:
\begin{equation}
     M_{j} = \frac{\Gamma_{j}}{1 + \Omega_{j}} \in [0, 1].
\label{eq: local magnitude score}
\end{equation}
This is computed at each dynamical point $\psi({\xi_{n}, \tau_{m}})$, i.e. for each column of $\mathbf{\hat{{\Lambda}}}$, where
\begin{equation}
     \Omega_{j} = \log_{10}[\max(\mathbf{s}_{j})] - \log_{10}[\min(\mathbf{s}_{j})] \in [0, \infty),
\label{eq: magnitude difference}
\end{equation}
\noindent and
\begin{equation}
    \Gamma_{j} =
    \begin{cases}
      \dfrac{\log_{10}[\min(\mathbf{s}_{j}) - \max(\mathbf{d}_{j})]}{\log_{10}[\min(\mathbf{s}_{j}) + \max(\mathbf{d}_{j})]}, & \text{if}\ \min(\mathbf{s}_{j}) > \max(\mathbf{d}_{j}) \\
      0, & \text{if}\ \min(\mathbf{s}_{j}) \leq \max(\mathbf{d}_{j})
    \end{cases}.
\label{eq: magnitude gap}
\end{equation}
\noindent Here $\Gamma_{j} \in [0,1]$ after imposing the floor condition $\Gamma_{j} = 0$ if $\Gamma_{j} < 0$ (see~\cite{Kaiser-2022} for details). It can be seen from the above that the proposed metric is based on the relative magnitudes of terms in some local region of the dynamics as opposed to the hard threshold that can be applied throughout the whole domain. This becomes especially relevant when dealing with the multiscale physical systems.

Finally, the procedure can be applied to all the GMM clusters, identified at the first step:
\begin{enumerate}
\item For each of the clusters we compute the $M$ score averaged over all the dynamical points that belong to the selected cluster and for all possible dominant balances (for each cluster we get $2^{K - 1} - 1$ averaged $M$ scores);
\item Each cluster is, then, assigned to the particular dominant balance that obtained the highest averaged $M$ score;
\item Clusters assigned to the same dominant balances are grouped together to form the final dominant balance models;
\item These dominant balance models can be directly mapped to the original $(\xi,\tau)$ domain and displayed using different color codes for the comparison with the standard evolution map.
\end{enumerate}

\section{Additional Examples of Dominant Balance Search}

In this section, we apply the dominant balance regime search to some additional representative examples of optical fibre propagation.

\subsection{Riemann wave propagation in the normal dispersion regime}

The first additional case that we consider considers propagation in the fiber normal dispersion regime, with a more complex example of dynamics than optical wavebreaking considered in the main manuscript.  Specifically, we again
consider the normalised NLSE: $i \, \psi_\xi - \psi_{\tau\tau} + |\psi|^2\psi = 0$, but we take initial conditions corresponding to a Riemann wave with a specific initial chirp profile that is proportional to the pulse amplitude. This is written here as:
\begin{equation}
\psi(0,\tau) = N\exp{(-\tau^{2}/4 - i \phi)},
\end{equation}
with
\begin{equation}
\phi = \sqrt{2} N \int_{-\infty}^{\tau} \exp{(-\tau^{\prime 2}/4)} \,d\tau^\prime = \sqrt{2 \pi}N [(1 + \erf(\tau/2)],
\end{equation}
\noindent where $\xi$, $\tau$, and $\psi$ represent the dimensionless coordinate, time and field, as introduced in the main manuscript and $N = \sqrt{L_{D}/L_{NL}} \approx 5.84$ was used in the simulations. With a Riemann wave initial condition, the propagation dynamics can be well-approximated by the inviscid Burger's equation, and are associated with pulse envelope steepening, and eventually the formation of a gradient catastrophe and characteristic shock formation \cite{Whitham-1999,Wetzel-2016}.

\begin{figure*}[ht]
\centering
\fbox{\includegraphics[width=\linewidth]{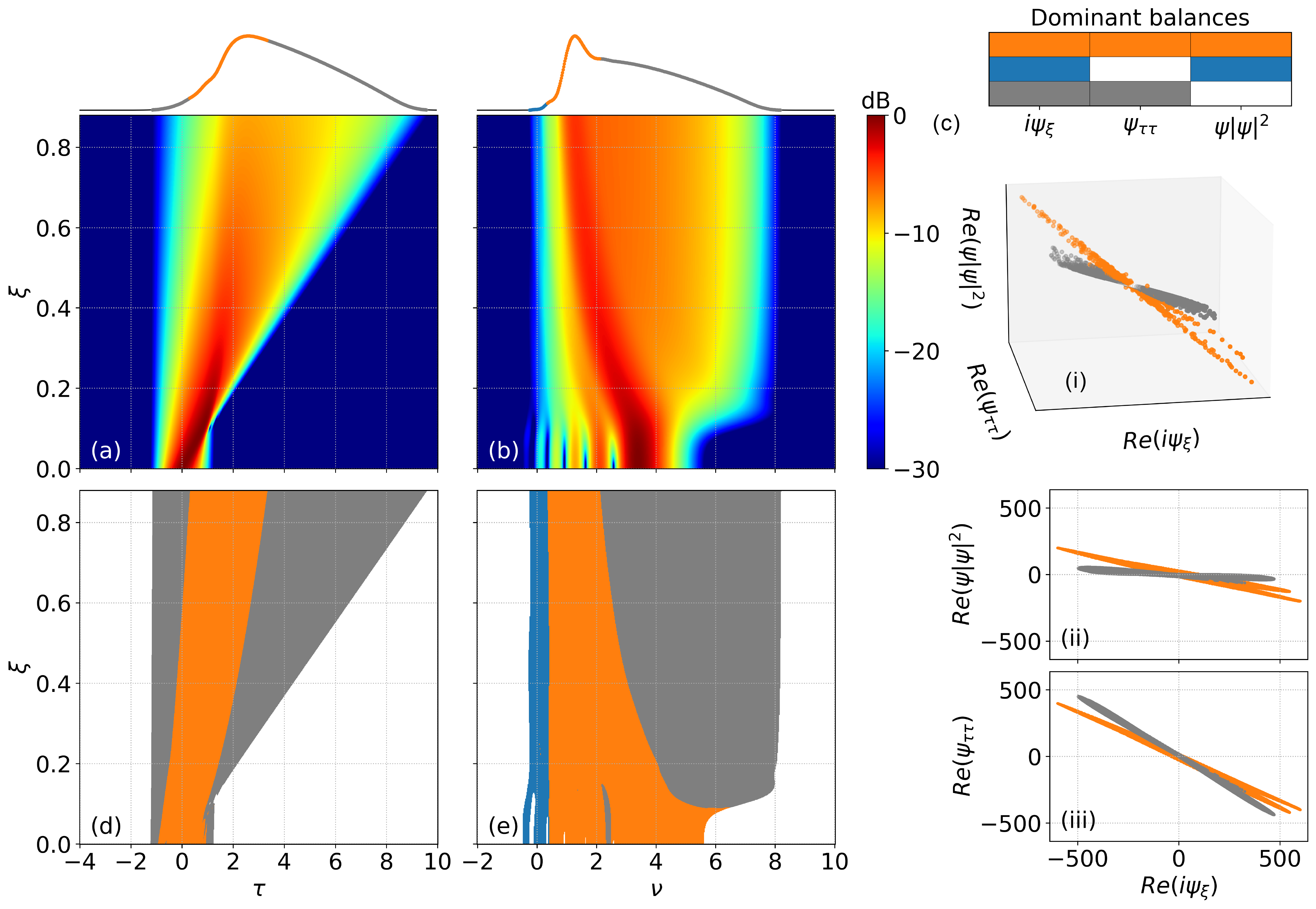}}
\caption{(a) temporal and (b) spectral intensity evolution maps for Riemann wave initial conditions. Both maps are normalised to a maximum intensity of unity and use the same logarithmic intensity scale. We also plot output temporal and spectral profiles on a linear scale. (c) plots equation space terms to illustrate the clustering  seen in the different projections. (d) and (e) plot the dominant balance regions for comparison with the intensity evolution maps.  The color code used in the spectral dominant balance plot corresponds to the Fourier transforms of the temporal terms shown in the legend.}
\end{figure*}

The results of simulation and dominant balance analysis for this case are shown in Fig.~3 which shows both evolution and dominant balance maps as in the main manuscript. The equation space clusters are very clearly delineated for this case, and we see only two clusters: one where involving both dispersion and nonlinearity, and another dominated only by dispersion. The evolution dynamics are associated with strong temporal steepening in the initial propagation (up to $\xi \sim 0.1$ after which the evolution of the steepened leading edge and central portion of the temporal profile remain governed by the full NLSE, with this sitting upon a broader dispersive background. This is contrast to the case of optical wavebreaking in the main manuscript (Fig.~1) where the initial evolution is dominated by the nonlinear term.
\newpage

\subsection{Perturbed fundamental soliton propagation}
A further instructive example is the well-known case of propagation of an optical soliton, with normalised NLSE: $i \, \psi_\xi + \psi_{\tau\tau} + |\psi|^2\psi = 0$. The sign change in the dispersive term compared to the NLSE used in the case of the Riemann wave above is because these dynamics are in the anomalous dispersion regime.  In addition, to illustrate the dynamics more generally, we show the results for an the initial condition of $\psi(0,\tau) = N \,\mathrm{sech}(\tau/\sqrt{2})$ where $N=1.5$. This will also allow us to see the effect of soliton perturbation. The results showing evolution and dominant balance maps are shown in Fig.~4.

\begin{figure*}[ht]
\centering
\fbox{\includegraphics[width=\linewidth]{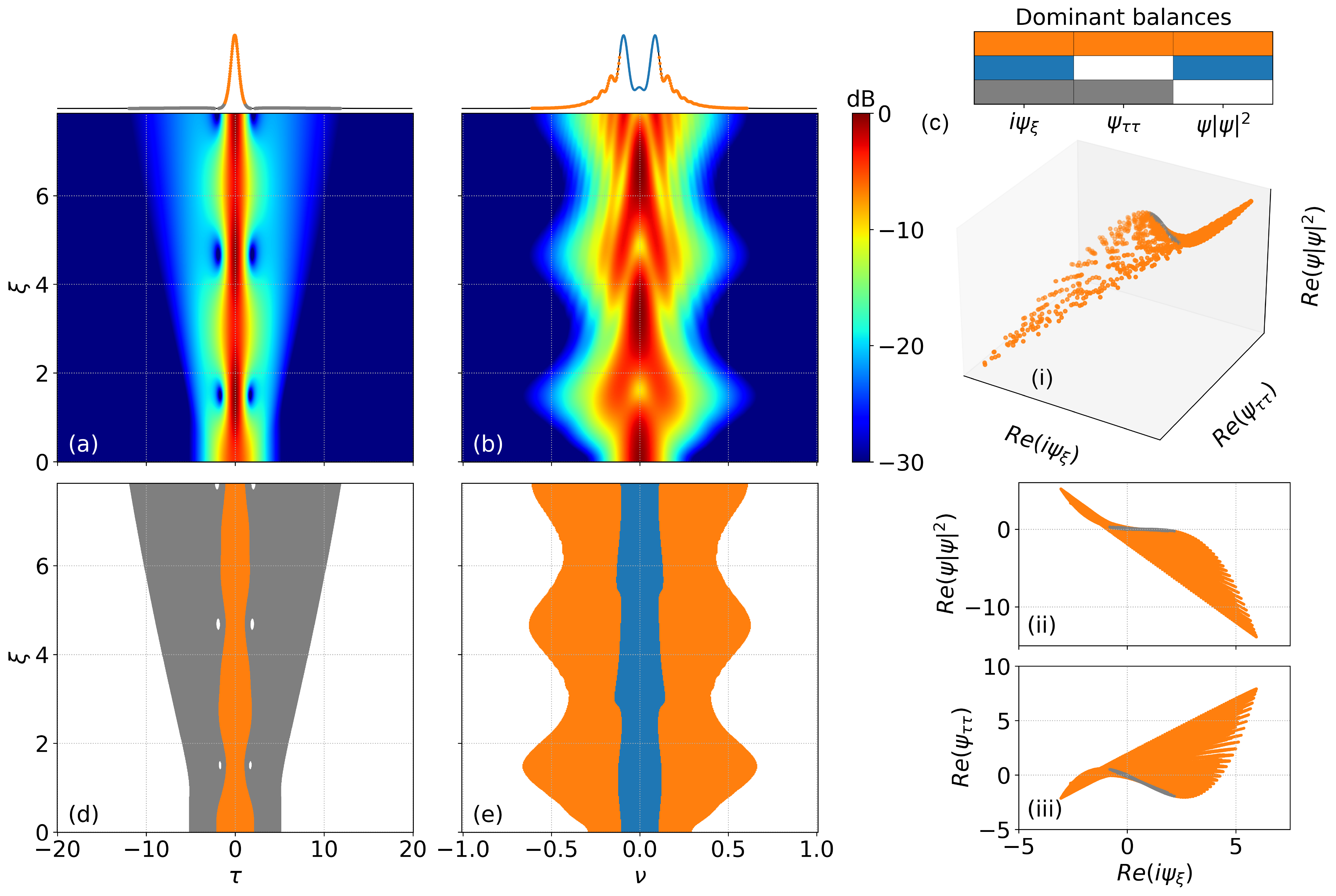}}
\caption{(a) temporal and (b) spectral intensity evolution maps for perturbed fundamental soliton dynamics with $N=1.5$. Both maps are normalised to a maximum intensity of unity and use the same logarithmic intensity scale. We also plot output temporal and spectral profiles on a linear scale. (c) plots equation space terms to illustrate the clustering  seen in the different projections. (d) and (e) plot the dominant balance regions for comparison with the intensity evolution maps.  The color code used in the spectral dominant balance plot corresponds to the Fourier transforms of the temporal terms shown in the legend.}
\end{figure*}

\newpage

For this case, the evolution plots in Figs~S1(a) and (b) show the expected dynamics for a perturbed soliton as it undergoes oscillatory temporal and spectral evolution, and shedding energy from the central temporal structure into  dispersive continuum radiation \cite{Agrawal-2019}.  The plots in Figs~S1(d) and (e) show how this is manifested in terms of dominant balance in both temporal and spectral domains respectively with the full NLSE driving evolution across the central temporal component and dispersion dominating the wings. In contrast, the role of isolated dispersion does not appear dominant in the frequency domain map; we see the centre of the spectrum is dominated only by nonlinearity (the NLSE dispersion operator in the frequency domain does not act on the spectral centre because of its $\nu^2$ dependence) whereas the wings of the spectrum show full NLSE contribution.

\subsection{Third-order dispersion-induced soliton fission}
Finally, we consider the case of higher-order soliton dynamics in the presence of only the third-order dispersion in the focusing regime $i \, \psi_\xi + \psi_{\tau\tau} + i \delta\psi_{\tau\tau\tau} + |\psi|^2\psi = 0$. The results show similar general features to the case of soliton fission considered in the main manuscript, but the absence of Raman scattering allows the role of dispersive perturbation to be examined separately.  The initial condition here correspond to a hyperbolic secant input pulse $\psi(0,\tau) = N \, \mathrm{sech}(\tau/\sqrt{2})$ with $N=3$. Third-order dispersion is included through the dimensionless parameter $\delta = \sqrt{2} \beta_{3}/ 3 T_{0} |\beta_{2}| \approx 0.06$ (comparable to the value in the main manuscript given that $f_{R} = 0$ in this case.)

Figures~5(a) and (b) plot the temporal and spectral evolution respectively, showing typical soliton fission characteristics. The temporal dominant balance map in Fig.~5(d) shows that the initial stage of the propagation is associated with temporal compression, and is primarily governed by the nonlinearity because of the higher value of $N$ associated with the injected higher-order soliton (dark blue).  After $\xi \sim 0.1$ the contribution of third-order dispersion becomes significant and drives the soliton fission at $\xi \sim 0.4$ accompanied by clear ejection of the dispersive wave. Here the red regions show comparable contribution from all terms across the temporal centre of the pulse, while the clear dominance of only the linear terms (light blue) is clear in the ejected dispersive wave properties in both the time and frequency domains.

\newpage

\begin{figure*}[ht]
\centering
\fbox{\includegraphics[width=\linewidth]{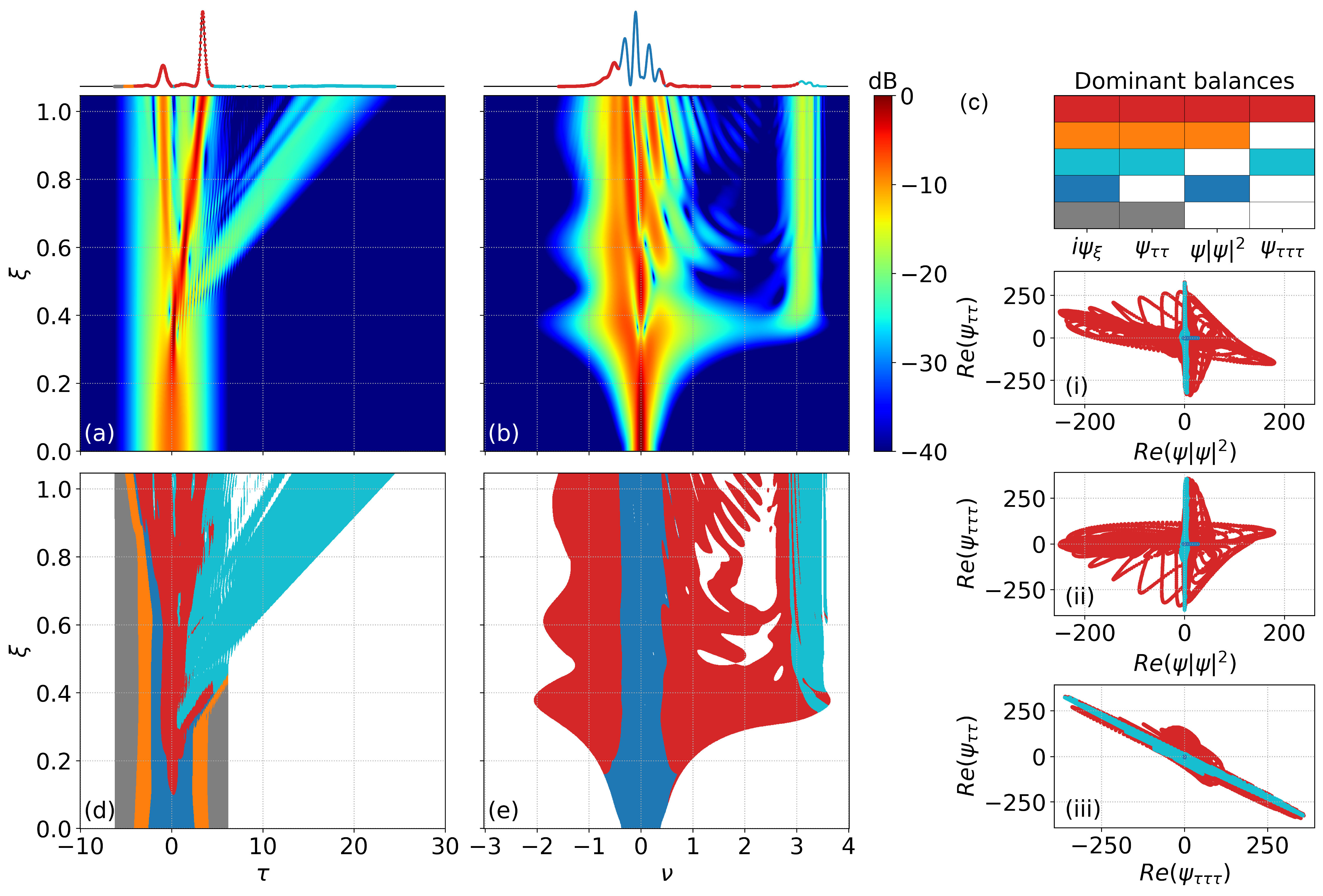}}
\caption{(a) temporal and (b) spectral intensity evolution maps for the soliton fission process induced by the third-order dispersion. Both maps are normalised to a maximum intensity of unity and use the same logarithmic intensity scale. We also plot output temporal and spectral profiles on a linear scale. (c) plots equation space terms to illustrate the clustering  seen in the different projections. (d) and (e) plot the dominant balance regions for comparison with the intensity evolution maps.  The color code used in the spectral dominant balance plot corresponds to the Fourier transforms of the temporal terms shown in the legend.}
\end{figure*}

\newpage

\bibliography{Ermolaev-2024}

\end{document}